\documentclass[12pt]{article} \textheight 8.5in \textwidth 6.5in
\title{Comment on Inflation and
Alternative Cosmology} \author{Stefan Hollands\thanks{\tt
stefan@gr.uchicago.edu} and Robert M. Wald\thanks{\tt
rmwa@midway.uchicago.edu} \\ \it Enrico Fermi Institute and Department
of Physics \\ \it University of Chicago \\ \it 5640 S.~Ellis Avenue,
Chicago, IL~60637, USA}
\begin{document}
\maketitle
\begin{abstract}

We respond to, and comment upon, a number of points raised in a recent
paper by Kofman, Linde, and Mukhanov.

\end{abstract}

\section{Introduction}

In a recent paper \cite{hw}, we argued that inflation does not provide
a satisfactory explanation of the ``special state'' (i.e.,
homogeneity, isotropy, and flatness) of our universe. We then noted
that the fundamental mechanism by which inflationary models produce
cosmologically significant departures from homogeneity and isotropy
involves the evolutionary behavior of modes whose proper wavelength is
larger than the Hubble radius. Since this evolutionary behavior will
occur whether or not inflation took place, we raised the possibility
that one could account for the departures from homogeneity and
isotropy in our universe via the same basic mechanism as in
inflationary models, but without postulating that the universe
actually underwent an era of inflation. We then identified a set of
assumptions concerning initial conditions that would be sufficient to
produce a perturbation spectrum of the same nature as that of inflationary
models. Finally, we provided an explicit model in which these
assumptions were made, and we carried out the calculations to show
that a perturbation spectrum similar to that of inflationary models
does, indeed, result.

Our arguments and conclusions have recently been criticized on a
variety of grounds by Kofman, Linde, and Mukhanov \cite{klm}. The main
purpose of the present paper is to respond to these criticisms and to
add some general comments on irreversibility and on the use of
probabilistic arguments.

\section{Pre-Planckian Initial Conditions}

Inflationary models have the feature that, at sufficiently early
times, the modes of cosmological interest had proper wavelength much
smaller than the Hubble radius. Therefore, it seems natural to
postulate that these modes were ``born'' in their ground state. In
addition, on account of adiabatic evolution during the era when their
wavelength is much smaller than the Hubble radius, the precise time at
which they were ``born'' should not be important. In fact, upon closer
inspection, these assumptions are seen to be far from obvious: In
inflationary models, the modes of cosmological interest actually had
wavelength very much smaller than the Planck scale at the onset of
inflation. Therefore, if sub-Planckian physics deviates substantially
from extrapolations of physics at scales larger than the Planck scale,
the validity of inflationary models would be in
doubt\footnote{However, investigations of some simple modifications of
sub-Planckian physics indicates that inflationary models are not very
sensitive to such modifications \cite{bm}.}. Nevertheless,
inflationary models provide a simple ``default assumption'' concerning
the initial state of modes of cosmological interest.

By contrast, in non-inflationary models, the modes of cosmological
interest have proper wavelength much larger than the Hubble radius
throughout the entire early evolutionary history of the
universe. Therefore, one must directly confront the issue of the
initial state of such modes. There is no theory presently capable of
``predicting'' the initial state of these modes, so one can only make
hypotheses and assumptions. In \cite{hw}, we put forward the
hypothesis that a semiclassical description of physics may be possible
at arbitrarily early times on length scales larger than some
fundamental scale $l_0$. As we explained in footnote 7 of \cite{hw},
we do {\em not} suggest that an accurate semiclassical description
would be obtained by a naive extrapolation of a semiclassical solution
to Einstein's equation to times earlier than the Planck time,
$t_P$. Rather, we proposed that some suitable ``coarse graining'' of
the degrees of freedom of quantum gravity over length scales smaller
than $l_0$ could yield an accurate (but, as yet, unknown to us)
semiclassical description of nature. We further proposed that when
modes of the quantum field ``emerge from the spacetime foam'' at
lengthscale $l_0$, they do so in their ground state.

Our above assumptions concerning the initial state of the
cosmologically relevant modes was criticized in \cite{klm} on the
grounds that they do not follow from quantum field theory in curved
spacetime. This criticism is, of course, entirely valid, since as
already noted above, at the present time there does not exist any
theory that is capable of predicting the initial state of these modes.
Our assumptions were also criticized in \cite{klm} on the grounds that
our assumed initial state depends only on $l_0$ whereas there are
other scales present---in particular, the Hubble radius---that are
dynamically more important in the evolution of these modes. This
criticism is a more significant one. The only scale that directly
enters the Lagrangian of a massless mode (see eq.~(4) of~\cite{hw}) is
the proper wavelength of that mode, which by our assumptions is equal
to $l_0$ at the semiclassical ``birth'' of the mode. Thus, we feel that our
assumption that the modes are ``born'' in their ground state is not
entirely unreasonable or unnatural---at least as an initial
hypothesis.  Nevertheless, we have no argument that other dynamically
relevant scales could not enter their initial state. Clearly, the
general issue of when modes can be treated semiclassically and in what
state they are ``born'' is a deep issue, about which very little is
known at present.

In \cite{hw}, we also presented an explicit model---involving
quantized sound waves of a perfect fluid---in which a perturbation
spectrum was obtained that is of the same nature as occurs in
inflationary models. As we stated very explicitly in the paper, we do
not expect this model to be a realistic description of nature in the
very early universe (particularly at times earlier than the Planck
time). Rather, the purpose of presenting this model was to simply to
illustrate how the ``overdamped'' evolution of modes with wavelength
larger than the Hubble radius could produce cosmologically relevant
perturbations in a context that is very different from inflationary
models---thereby providing an ``existence proof'' of alternative
models to inflationary ones. This model was sharply criticized in
\cite{klm} for its unrealistic properties. We, of course, agree that
the model we presented is not realistic, but this does not mean that
there could not exist more realistic models that share its basic
features.

The criticisms of \cite{klm} very nicely emphasize the point (which we
also made in our paper) that in non-inflationary models, the time at
which the cosmologically relevant modes have proper wavelength equal
to $l_0$ occurs at times much earlier than the Planck time (or, more
precisely, in a region of spacetime corresponding to times much
earlier than the Planck time in naive extrapolations of classical or
semiclassical models). Consequently, it is not likely that any
reliable calculations can be done for non-inflationary models until
our understanding of quantum gravitational physics improves
significantly. However, we strongly disagree with the assertion of
\cite{klm} that this fact provides an additional argument in favor of
inflation. More precisely, it provides an argument in favor of
inflation only in the sense that the (true) statement ``If you are
going to find your keys tonight, then you must have dropped them under
the lamppost'' provides an argument that you actually dropped your
keys under the lamppost. We believe that the keys to understanding the
origin of structure in our universe are likely to lie in the
pre-Planckian era.

\section{Irreversible Processes and the Second Law of Thermodynamics}

It is clear that the present state of our universe is very ``special''
in the sense that its entropy is very much less than that of corresponding
universes that are similar on large (i.e., Hubble) scales but are much
more gravitationally clumped on small (e.g., galactic) scales. It is
this ``specialness'' of our present universe---i.e., the fact that its
entropy is very far from its maximum possible value---that gives rise
to the second law of thermodynamics. In \cite{hw}, we
argued---following arguments previously given in \cite{penrose}---that rather
than seeking to use the second law of thermodynamics or other
dynamical arguments to explain how the universe arrived at its current
state starting from arbitrary initial conditions, we should be seeking
to use the (as yet to be developed) theory of initial conditions of
the universe to explain how the second law of thermodynamics came into
being. We argued further that inflationary models do not avoid the
necessity of assuming ``special'' initial conditions for our
universe. Indeed, we argued that if an expanding universe generically
undergoes an era of inflation, then---unless one introduces
assumptions that break time reversal invariance---a collapsing
universe must generically undergo an era of ``deflation''. We view
this result as a reducto ad absurdum for the claim that special initial
conditions are not needed for inflation to occur.

Our arguments and conclusions were criticized in \cite{klm} on a
number of grounds. In particular, the simple model of a homogeneous
scalar field in a flat Robertson-Walker universe was analyzed, and it
was claimed that---in contradiction with our conclusions---in this
model, inflation is an ``attractor'' whereas deflation is not. The
analysis of this model given in \cite{klm} relies heavily on
assumptions concerning the probability measure on initial conditions,
so we will defer our discussion of this model to the next section. In
the present section, we will comment on some of the other arguments
concerning irreversibility given in \cite{klm}.

We agree with the claims of \cite{klm} that many ``irreversible
processes'' have occurred in the evolutionary history of our universe;
indeed, the fact that irreversible processes have occurred is the main
basis of our claim that the initial entropy of our universe was very
low, i.e., that the universe began in a very ``special'' initial
state. We also agree with their assertion that if one time-reversed [a
very slight modification of] the initial conditions representing the
present state of our universe, the subsequent evolution would not
produce anything similar to the (time reverse of) the initial
conditions that our universe started with. Indeed, the initial state
of our universe appears to have been quite ``smooth'', whereas the
generic final state of a collapsing universe would be expected to be
extremely ``messy'', with numerous black holes forming and merging,
etc. In particular, we certainly would not expect an era of deflation to
occur just prior to the ``big crunch''. But this is just our point:
Why should not the initial state of the universe have been
correspondingly messy? Why should there not have been many regions of
``delayed big bang singularities'' (i.e., white holes) filling the
early universe, and, indeed, filling the present universe as well? We
believe that the answer to these questions is that the initial state
of the universe was very ``special''.

We strongly disagree with the claim in \cite{klm} that particle
production changes the number of degrees of freedom of a
system\footnote{We note however that, while particle production does
not change the number of degrees of freedom, the expansion of the
universe {\em does} produce an effective change in the number of
semiclassical degrees of freedom in the universe, since new
semiclassical modes presumably ``emerge from the spacetime foam'' as
the universe expands. In both inflationary models and in our proposal,
these modes are assumed to emerge in their ground state and, hence,
their emergence is not associated with particle creation. However, it
is possible that this phenomenon may play a key role in accounting for
the thermodynamic arrow of time.}. Fundamentally, in quantum field
theory the degrees of freedom reside in the field and are present
whether or not the modes of the field are in their ground state or in
excited states. The process of particle creation in the early universe
is irreversible only in the same sense as the breaking of an egg, but
not in any more fundamental sense. Contrary to the claim
in~\cite{klm}, such ``irreversible'' dynamics in classical statistical
physics is associated with a measure preserving flow on phase space.

\section{The Probability of Inflation}

We begin this section with two completely general remarks concerning
the use of probabilistic arguments. First, probabilistic arguments can
be used reliably when one completely understands both the nature of
the underlying dynamics of the system and the source of its
``randomness''. Thus, for example, probabilistic arguments are very
successful in predicting the (likely) outcomes of a series of coin
tosses. Conversely, probabilistic arguments are notoriously unreliable
when one does not understand the underlying nature of the system
and/or the source of its randomness. For example, if asked to estimate
the conditional probability that if cows were to be discovered on a
distant planet, then their color would be green, a physicist might
proceed by taking the frequency bandwidth of the green part of the
visible spectrum and dividing this by the bandwidth of the entire
visible spectrum. By contrast, a chemist might take the number of
common chemical compounds that are green and then divide this quantity
by the total number of common chemical compounds. On the other hand, a
biologist might first estimate the likelihood of the color of the
planet itself by methods similar to that of the chemist, and then take
camouflage and other survival factors into account to estimate the
probability of the cow being green. Although each of these methods of
estimation is arguably reasonable, it is not likely that any of them is
reliable, particularly if the planet and the cows living on it are
very different from anything we have experienced.

The second comment concerns the general situation where one has a
manifold $M$ representing the possible states of a system, on which
there is defined a measure $\mu$ such that $\mu(M) = \infty$. Consider
a property, $Q$, of the system that corresponds to a (measureable)
subset, $S$, of $M$. Suppose that we wish to know the probability,
$p(Q)$, that property $Q$ holds. Then there are precisely three
possibilities:

\begin{itemize}

\item $\mu(S) < \infty$. In this case, $p(Q) = 0$.

\item $\mu(M - S) < \infty$. In this case $p(Q) = 1$.

\item $\mu(S) = \infty$ and $\mu(M - S) = \infty$. In this case $p(Q)$
is undefined.

\end{itemize}
In the last case, one might be tempted to define $p(Q)$ by considering
a nested family of sets, $K_n$, of finite measure whose union is $M$,
and considering the limit as $n \rightarrow \infty$ of $\mu(S \cap
K_n)/\mu(K_n)$. In practice, this might be done by introducing
coordinates on $M$ and taking $K_n$ to be the coordinate ball of
radius $n$. However, it is easy to see that one can get any answer for
$p(Q)$ that one wishes by making a suitable choice of the sets $K_n$ or,
equivalently, by a suitable choice of coordinates on $M$.

In \cite{klm}, the model of a homogeneous scalar field in a spatially
flat ($k=0$) Robertson-Walker universe was considered, and it was
claimed that this model provides a counterexample to our claim that
the probability of inflation should equal the probability of
deflation. In fact, this model---including its generalization to the
$k=\pm 1$ cases---was previously analyzed in detail by Hawking and
Page \cite{hp}. The $k=+1$ case is particularly relevant, since in
this case all (or essentially all) solutions should start from a ``big
bang'' singularity and end in a ``big crunch'' singularity, so the
probabilities of inflation and deflation can be compared.

The model of \cite{klm} and \cite{hp} is a constrained Hamiltonian
system on a 4-dimensional phase space, $M$. We choose as coordinates
on $M$ the Robertson-Walker scale factor, $a$, the value of the scalar
field, $\phi$, and the canonically conjugate momenta, $p_a = - a
\dot{a}$ and $p_\phi = a^3 \dot{\phi}$, respectively, where the
overdot denotes the derivative with respect to proper time. The
Hamiltonian of this system is given by \cite{hp}
\begin{equation}
{\mathcal H} = \frac{1}{2}(-a^{-1} p_a^2 + a^{-3} p_\phi^2 - ka + m^2
a^3 \phi^2)
\end{equation}
and the constraint hypersurface, $\mathcal C$, is given by ${\mathcal
H} = 0$. 

As is well known, the Liouville measure, defined by the volume element
\begin{equation}
{}^{(4)}\epsilon =  dp_a \wedge da \wedge dp_\phi \wedge d\phi,
\end{equation}
is invariant under dynamical evolution. It seems to be much less widely
known that this measure induces a natural invariant measure, $\mu$, on
on the constraint hypersurface, $\mathcal C$. This measure is given
by the volume element ${}^{(3)}\epsilon$ on $\mathcal C$ that is
determined by the condition that on $\mathcal C$ we have
\begin{equation}
d{\mathcal H} \wedge {}^{(3)}\epsilon = {}^{(4)}\epsilon
\end{equation}
(see, e.g., section 7 of \cite{k}). Consequently, one way of defining
the ``probability of inflation'', $p(I)$, that appears to be very
natural at least from a mathematical point of view would be to set
$p(I) = \mu({\mathcal I})/\mu({\mathcal C})$, where $\mathcal I$
denotes the region of $\mathcal C$ that is occupied by dynamical
orbits that undergo an era of inflation. It is obvious that with this
definition, we have $p(I) = p(D)$, where $p(D)$ denotes the
similarly defined probability of deflation. However, by an analysis
similar to that given in \cite{hp}, it is not difficult to show that
$\mu({\mathcal C}) = \infty$, $\mu({\mathcal I}) = \infty$, and
$\mu({\mathcal C} - {\mathcal I}) = \infty$, so, by our general remarks
above, the probabilities of inflation and deflation are not defined.

Alternatively, following \cite{hp}, one could choose an initial data
surface, $\Sigma$ on $\mathcal C$---i.e., a two-dimensional surface on
$\mathcal C$ that is intersected transversely by all (or almost all)
of the dynamical orbits on $\mathcal C$---and try to define a measure
directly on $\Sigma$. As noted in \cite{hp}, the pullback of the
symplectic form, $\Omega = dp_a \wedge da + dp_\phi \wedge d\phi$, to
$\Sigma$ yields a mathematically natural volume element on $\Sigma$
that is invariant under dynamical evolution. One could then set $p'(I)
= \mu'({\mathcal I})/\mu'(\Sigma)$ where $\mu'$ denotes the measure on
$\Sigma$ associated with the pullback of $\Omega$ to $\Sigma$, and
${\mathcal I}$ now denotes the region of $\Sigma$ corresponding to
initial data for inflationary solutions. In contrast to the
probability $p(I)$ defined in the previous paragraph, the probability
$p'(I)$ will depend upon the choice of initial data surface
$\Sigma$---although it will be invariant under dynamical evolution of
$\Sigma$. For a choice of initial data surface of the form $f(\phi) =
\rm{const.}$, it was shown in \cite{hp} that $\mu'(\Sigma) = \infty$,
$\mu'({\mathcal I}) = \infty$, and $\mu'(\Sigma - {\mathcal I}) =
\infty$. Thus, for the measure $\mu'$, as with the measure $\mu$ above,
the probability of inflation is not defined.  Indeed, section 5 of
\cite{hp} provides a nice illustration of the above general point that
if one attempts to define $p'(I)$ by a limiting procedure, the answer
one obtains will depend upon how the limit is taken.

Alternatively, following \cite{klm}, if one restricts attention to the
case $k = 0$, then---after taking the constraint ${\mathcal H} = 0$
into account---the dynamical evolution equations for $\phi$ can be
expressed in terms of $\phi$ and $\dot{\phi}$ alone, i.e., they do not
depend upon $a$. One may then (arbitrarily) ignore the coordinate $a$
on the constraint hypersurface $\mathcal C$ and work in a new two
dimensional space $\tilde{\mathcal C}$ parametrized by $\phi$ and
$\dot{\phi}$.  (This cannot be done for the cases $k = \pm 1$ since
the dynamics of $\phi$ in these cases depends on $a$, so this
procedure is very special to the case $k = 0$.) One may then
(arbitrarily) define a Euclidean metric, $g$, on $\tilde{\mathcal C}$
by treating $\phi$ and $\dot{\phi}$ as though they were Cartesian
coordinates. If one then chooses an initial data surface
$\tilde{\Sigma}$ on $\tilde{\mathcal C}$, this Euclidean metric $g$
will induce a Riemannian metric $h$ on $\tilde{\Sigma}$ and, thereby,
a measure, $\tilde{\mu}$, on $\tilde{\Sigma}$. This measure will not
be preserved under dynamical evolution. One could then set
$\tilde{p}(I) = \tilde{\mu}({\mathcal I})/\tilde{\mu}(\tilde{\Sigma})$
where ${\mathcal I}$ now denotes the region of $\tilde{\Sigma}$
corresponding to initial data for inflationary solutions. This is
precisely what was done in \cite{klm}, with the initial data surface
chosen to be $H = l_P^{-1}$, where $H = \dot{a}/a$ denotes the Hubble
constant (which is related to $\phi$ and $\dot{\phi}$ by the
constraint). In this case, $\tilde{\mu}(\tilde{\Sigma}) < \infty$, so
a well defined result for $\tilde{p}(I)$ is obtained. The numerical
investigations of \cite{klm} establish that $\tilde{p}(I)$ is nearly
$1$ if $m \ll m_P$ (with most of the initial data giving rise to
inflation satisfying $\phi \gg m_P$). However, since the measure
used in \cite{klm} is not preserved under dynamical evolution, a
different answer will be obtained if one uses the same procedure to
calculate probabilities on a time evolved initial data surface. It was
shown in \cite{klm} that a very small value would be obtained for the
probability, $\tilde{p}(I)$, that inflation occurred in the early
universe if one were to do the corresponding calculations using a late
time initial data surface. This was interpreted in \cite{klm} as
showing that the probability for deflation is small.

Clearly, many other possible choices of measure could be made and,
correspondingly, many other conclusions of all varieties could be
drawn. Indeed, the situation becomes even more murky when one
considers more general, inhomogeneous models. In our view, the
probability of inflation cannot be reliably estimated until one has
some understanding of the physical processes that determine the
initial state of our universe.

This research was supported in part by NSF grant PHY00-90138 to the
University of Chicago.

\end{document}